\documentclass[11pt,a4paper]{article}
\usepackage{here}
\usepackage{eurosym}
\usepackage{graphicx}
\usepackage{amsmath}
\usepackage{amssymb}
\usepackage{latexsym}
\usepackage{color}
\usepackage{epstopdf}
\usepackage{subfigure}
\usepackage[utf8]{inputenc}
\usepackage{graphicx}
\usepackage{braket}
\usepackage[french,english]{babel}
\usepackage[T1]{fontenc}
\usepackage{float}
\usepackage{cite}
\usepackage{filecontents}
\usepackage[colorlinks={true}]{hyperref}
\hypersetup{colorlinks=true,linkcolor=red,citecolor=magenta,urlcolor=blue}
\usepackage[compact]{titlesec}

\usepackage{xcolor}
\usepackage{color}
\usepackage{hyperref}
\usepackage{cleveref}
\usepackage{lipsum}
\usepackage{physics}
\usepackage{dsfont}
\usepackage[switch]{lineno}
\usepackage[utf8]{inputenc}
\usepackage{bm}

\usepackage[font=footnotesize,labelfont=bf]{caption}

\usepackage{orcidlink}
\usepackage[hmargin=2cm,vmargin=2cm,rmargin=2cm, lmargin=2cm]{geometry}
\hypersetup{colorlinks=true,citecolor=red,filecolor=blue,linkcolor=blue,urlcolor=blue}
\DeclareUnicodeCharacter{FB01}{fi}
\DeclareUnicodeCharacter{200B}{}

\def\keywordname{{\bfseries \emph{Keywords}}}%
\def\keywords#1{\par\addvspace\medskipamount{\rightskip=0pt plus1cm
\def\and{\ifhmode\unskip\nobreak\fi\ $\cdot$
}\noindent\keywordname\enspace\ignorespaces#1\par}}

\begin{document}
\renewcommand{\figurename}{\textbf{Fig.}}
\renewcommand{\tablename}{\textbf{Table}}
\newcommand{\un}{\mathds{1}}
\newcommand{\1}{\mathbbm{1}}
\newcommand{\vect}[1]{\boldsymbol{#1}}

\newcommand{\era}{\end{array}}
\newcommand{\beq}{\begin{equation}}
\newcommand{\eeq}{\end{equation}}
\newcommand{\beqar}{\begin{eqnarray}}
\newcommand{\eeqar}{\end{eqnarray}}

\newcommand{\lb}{\label}
\thispagestyle{empty}

\baselineskip=18pt
\medskip

\begin{center}
~~~~~~~~~~~~~~~
\\
\vspace{2cm}
\noindent { {\textbf{Dimeric perylene-bisimide organic molecules: Application as a quantum battery}}}\\
\vspace{0.7cm}
\noindent
\vspace{0.5cm}
{\small Samira Elghaayda~\!\!\orcidlink{0000-0002-6655-0465}} $^{a,}${\footnote{E-mail: \textsf{\href{mailto:samira.elghaayda-etu@etu.univh2c.ma}{samira.elghaayda-etu@etu.univh2c.ma}}}} and {\small Mostafa Mansour~\!\!\orcidlink{0000-0003-0821-0582}} $^{a, }${\footnote {E-mail: \textsf{\href{mailto:mostafa.mansour.fsac@gmail.com}{mostafa.mansour.fsac@gmail.com}}}}

\noindent $^{a}${{\footnotesize Laboratory of Mechanics and High Energy Physics, Depart of Physics,\\ Faculty of Sciences of Aïn Chock, Hassan II University,\\  Casablanca , Morocco }}\\[0.5em]

\end{center}
\vspace{2cm}

\begin{abstract}
This work introduces a unified theoretical framework for quantum batteries (QBs) constructed from thermally equilibrated arrays of dimeric perylene bisimide (PBI) molecules. These organic dimers, with chemically tunable transition energies and dipole-dipole interactions, constitute a scalable and practical platform for quantum energy storage. Using exact diagonalization of the Gibbs state supported by analytic and numerical resource-theoretic tools, we evaluate four performance metrics: ergotropy, instantaneous charging power, storage capacity, and quantum coherence. We find that exact resonance ($\nu_1 = \nu_2$) suppresses both ergotropy and charging power due to symmetric thermal population distributions. Introducing finite detuning ($\Delta = \nu_1 - \nu_2$) breaks this symmetry, redistributes populations, and significantly enhances extractable work, charging power, and storage capacity. Furthermore, while the capacity remains invariant under unitary dynamics, providing a useful reference bound, intermediate dipole-dipole coupling strengths ($V_{12}$) optimize the trade-off between ergotropy, coherence retention, and storage performance. Crucially, coherence-assisted energy storage persists up to experimentally relevant temperatures, underscoring the thermal resilience of PBI-based QBs. These results establish spectral detuning and dipole-dipole interaction tuning as essential design principles, positioning PBI dimers as a chemically realistic, experimentally accessible, and thermodynamically robust platform that bridges molecular engineering with quantum energy storage.

\end{abstract}

\vspace{0.5cm}
\newpage

\section{Introduction\label{sec1}}
Quantum batteries (QBs) represent a frontier in energy storage technologies, exploiting quantum coherence, entanglement, and collective effects to outperform their classical analogs in speed and efficiency \cite{alicki2013entanglement, ferraro2018high,campaioli2024colloquium,quach2023quantum,ahmadi2024nonreciprocal}. A QB typically consists of one or more two-level quantum systems, capable of storing energy and releasing it through unitary or open-system dynamics \cite{binder2015quantacell, Andolina2019}. Theoretical studies have shown that under suitable protocols, such as superabsorption or dark-state stabilization, both charging power and stored energy can scale superextensively with system size \cite{quach2020using, shi2022entanglement, li2025collective}.  However, most protocols depend on finely tuned initial states \cite{kamin2020entanglement,shi2022entanglement}, and scalability is restricted by decoherence in large-scale quantum systems \cite{andolina2018charger,rossini2019many,yao2022optimal,dou2022cavity,PhysRevE.106.054107,mojaveri2024extracting,ali2024ergotropy,quach2020using,PhysRevLett.118.150601}. These challenges underscore the need for platforms that can sustain quantum resources under realistic conditions. Superconducting qubits \cite{HaddadiQB2024} are among the leading candidates for QB implementations, but they demand cryogenic temperatures and precise energy matching, which limits scalability and robustness. Recent advances in superconducting QBs have shown that careful tuning of Josephson energies and inter-qubit coupling can significantly enhance ergotropy, charging power, and capacity, with quantum coherence playing a dual role as both a resource for rapid charging and a stabilizing factor against energy oscillations \cite{elghaayda2025}. To overcome the constraints of cryogenic architectures, alternative platforms that preserve coherence and performance under ambient conditions are actively sought.

Organic molecular systems, particularly dimeric perylene–bisimide (PBI) arrays, present a promising platform for quantum energy storage. Their engineered dipole–dipole interactions ($V_{12}$) and tunable transition frequencies ($\nu_1$, $\nu_2$) support long-lived quantum coherence \cite{lang2007comparison,issac2014stepwise,issac2012single}, while chemical synthesis allows scalable fabrication without the constraints of cryogenic infrastructure \cite{bredas2017photovoltaic,polman2016photovoltaic,thorwart2009enhanced}. Quantum interference and coherence phenomena, such as coherent population trapping and electromagnetically induced transparency, have been shown to play a central role in controlling decoherence and enhancing quantum system performance \cite{ficek2005quantum}. In the context of molecular dimers, these interference mechanisms can be exploited to optimize energy storage and extraction, enabling faster generation and revival of coherent superpositions. Notably, PBI monomers exhibit dephasing times much longer than the gating times, allowing many coherent operations within a single coherence window \cite{reina2018conditional}. Furthermore, conditional quantum dynamics and nonlocal correlations in dimeric and trimeric molecular arrays have been demonstrated to produce entanglement and support quantum logic operations \cite{reina2018conditional}. These results suggest that molecular arrays with intermediate dipole–dipole coupling strengths can achieve an optimal balance between coherence and extractable work, consistent with our findings that intermediate $V_{12}$ maximizes ergotropy and charging efficiency. For molecular aggregates coupled to photonic modes, neither the zero‐interaction limit nor the ultra-strong coupling regime is optimal; instead, an intermediate exciton–exciton interaction strength enhances both charging power and stored energy density \cite{li2025collective}. Recent advances in single-molecule spectroscopy now permit picosecond-level control of PBI dimers \cite{basche2008single,weigel2015shaped,brinks2014ultrafast,accanto2017rapid,hildner2011femtosecond,ernst2009photoblinking}, positioning them as realistic candidates for coherence-assisted energy storage.

Our study builds on previous work on organic and collective QBs \cite{li2025collective,campaioli2024colloquium} by considering molecular dimers initialized in the global Gibbs state, which captures realistic thermal equilibrium conditions. We examine how temperature ($T$), spectral detuning ($\Delta = \nu_1 - \nu_2$),   the average transition frequency ($\nu_{0} $) and dipole–dipole coupling ($V_{12}$) jointly influence ergotropy, charging power, coherence, and storage in PBI dimers. In particular, spectral detuning enhances extractable work and charging efficiency, while intermediate $V_{12}$ balances coherence, ergotropy and capacity, optimizing performance without destroying quantum resources. By analyzing the Gibbs state of interacting dimers, we show that storage capacity is invariant under unitary dynamics but depends sensitively on the initial spectral structure and coupling strength. Importantly, coherence-driven energy storage persists at elevated temperatures, highlighting the thermodynamic resilience of PBI-based architectures. These results, supported by numerical simulations, provide insights into how molecular parameters can be tuned to maximize energy storage and extraction, complementing earlier theoretical and experimental studies and suggesting strategies for scalable, realistic QBs. 

The plan of the paper is as follows. Section \ref{sec2} introduces the model Hamiltonian and derives the thermal equilibrium state of PBI dimers. In Section \ref{sec3}, we define the key performance metrics: ergotropy, charging power, capacity, and coherence. Section \ref{sec4} examines the dynamical behavior of these metrics as functions of temperature $T$, detuning $\Delta=\nu_1-\nu_2$, dipole–dipole coupling $V_{12}$, and the average transition frequency $\nu_0=\tfrac{\nu_1+\nu_2}{2}$, while also exploring optimization strategies based on molecular design and tunable dipolar interactions. Finally, Section \ref{sec5} discusses the concluding remarks and the future research avenues toward the realization of scalable organic QBs. 

\section{Model and charging framework\label{sec2} }
\begin{figure}[ht]
    \centering
    \includegraphics[width=0.65\linewidth]{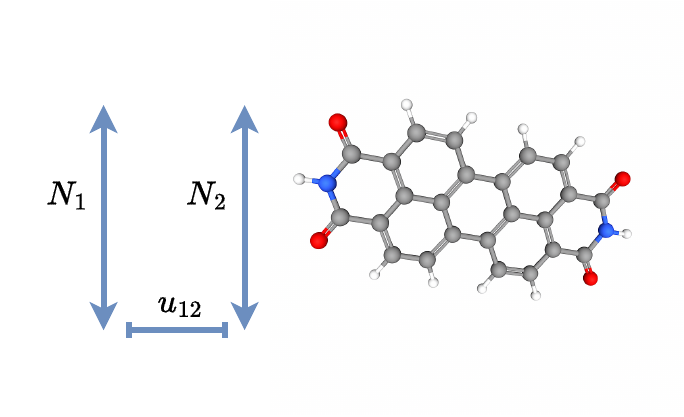}
    \caption{(Left) Schematic representation of PBI dimer, illustrating the transition dipole moments $\bm{N}_1$ and $\bm{N}_2$ separated by the intermolecular displacement vector $\bm{u}_{12}$. (Right) Molecular structure of the PBI organic molecule.}
    \label{fig01}
\end{figure}

This study investigates a molecular QB comprising a two-level dimer of PBI molecules, covalently linked via a rigid calix[4]arene bridge \cite{issac2014stepwise, issac2012single,hettich2002nanometer}. The use of PBI dimers is motivated by their well-characterized photophysics at the single-molecule level \cite{lang2007comparison}, including their robust dipole–dipole interactions and coherence lifetimes under ambient conditions. Recent advances in femtosecond-scale gate operations \cite{susa2010nonlocal} and radiative control in molecular systems \cite{reina2004radiative} further support their feasibility as platforms for room-temperature QBs. We model the two-level dimer as two identical quantum emitters interacting via dipole–dipole coupling. Each emitter is characterized by a transition frequency $\nu_i$ and a transition dipole moment $\hat{\bm{N}}_i \equiv \langle 0_i | \mathbf{D}_i | 1_i \rangle$, where $\mathbf{D}_i$ is the dipole operator. The emitters are separated by a vector $\bm{u}_{12}$, and experience spontaneous emission at rates $\Lambda_i$. The system configuration is illustrated in Fig. \ref{fig01}.

Following Refs. \cite{ficek2005quantum, reina2004radiative}, the coherent coupling strength $V_{12}$ and collective decay rate $\Lambda_{12}$ between the two emitters are given by:
\begin{align}
    \Lambda_{12} &= \frac{3}{2} \sqrt{\Lambda_1 \Lambda_2} \Bigg[ \left(\hat{\bm{N}}_1 \cdot \hat{\bm{N}}_2 - (\hat{\bm{N}}_1 \cdot \hat{\bm{u}}_{12})(\hat{\bm{N}}_2 \cdot \hat{\bm{u}}_{12}) \right) \frac{\sin z_{12}}{z_{12}} \nonumber \\
    &\quad + \left(\hat{\bm{N}}_1 \cdot \hat{\bm{N}}_2 - 3(\hat{\bm{N}}_1 \cdot \hat{\bm{u}}_{12})(\hat{\bm{N}}_2 \cdot \hat{\bm{u}}_{12}) \right) \left(\frac{\cos z_{12}}{z_{12}^2} - \frac{\sin z_{12}}{z_{12}^3}\right) \Bigg], \label{eq:gamma} \\
    V_{12} &= \frac{3}{4} \sqrt{\Lambda_1 \Lambda_2} \Bigg[ \left((\hat{\bm{N}}_1 \cdot \hat{\bm{u}}_{12})(\hat{\bm{N}}_2 \cdot \hat{\bm{u}}_{12}) - \hat{\bm{N}}_1 \cdot \hat{\bm{N}}_2 \right) \frac{\cos z_{12}}{z_{12}} \nonumber \\
    &\quad + \left(\hat{\bm{N}}_1 \cdot \hat{\bm{N}}_2 - 3(\hat{\bm{N}}_1 \cdot \hat{\bm{u}}_{12})(\hat{\bm{N}}_2 \cdot \hat{\bm{u}}_{12}) \right) \left(\frac{\cos z_{12}}{z_{12}^3} + \frac{\sin z_{12}}{z_{12}^2}\right) \Bigg], \label{eq:V12}
\end{align}
where $z_{12} = n k_{12} r_{12}$, $k_{12} = \omega_{12} / c$, $\omega_{12} = \pi (\nu_1 + \nu_2)$, and $n$ is the refractive index of the surrounding medium. The single-emitter spontaneous emission rate is $\Lambda_i = n \frac{\omega_i^3 \|\bm{N}_i\|^2}{3 \epsilon_0 \hbar c^3}$, scaling cubically with transition frequency $\omega_i$. For identical emitters with parallel dipoles oriented perpendicular to the separation axis, the coupling terms simplify to:
\begin{align}
    \Lambda_{12} &= \sqrt{\Lambda_1 \Lambda_2}, \label{eq:gamma_simple} \\
    V_{12} &= \frac{3 \sqrt{\Lambda_1 \Lambda_2}}{8 \pi z_{12}^3} \left[1 - 3 (\hat{\bm{N}}_1 \cdot \hat{\bm{u}}_{12})^2 \right]. \label{eq:V12_simple}
\end{align}
Under the Born–Markov approximation \cite{reina2014extracting, reina2018conditional,susa2012correlations}, the total Hamiltonian of the system reads as setting ($ h=1$):
\begin{equation}
    H_{\text{dimer}} = H_Q + H_{12}, \label{eq:H_dimer}
\end{equation}
where $H_Q = -\frac{1}{2} (\nu_1 \sigma_z^{(1)} + \nu_2 \sigma_z^{(2)})$ describes the local energy levels, and $H_{12} = \frac{1}{2} V_{12} (\sigma_x^{(1)} \otimes \sigma_x^{(2)} + \sigma_y^{(1)} \otimes \sigma_y^{(2)})$ captures the d–d interaction. Expressed in the standard basis $\{ |00\rangle, |01\rangle, |10\rangle, |11\rangle \}$, the matrix form is \cite{chouiba2025unveiling}:
\[
H_{\text{dimer}} =
\begin{pmatrix}
-\nu_0 & 0 & 0 & 0 \\
0 & -\frac{\Delta}{2} & V_{12} & 0 \\
0 & V_{12} & \frac{\Delta}{2} & 0 \\
0 & 0 & 0 & \nu_0
\end{pmatrix},
\]
where $\nu_0 = \frac{\nu_1 + \nu_2}{2}$ is the mean transition frequency, and $\Delta = \nu_1 - \nu_2$ is the detuning. The eigenvalues of the Hamiltonian $H_{\text{dimer}}$ and the corresponding eigenstates are:
\begin{align}
    \epsilon_1 &= - \nu_0, \quad &|\Phi_1\rangle &= |00\rangle, \nonumber \\
    \epsilon_2 &= -\frac{\alpha}{2}, \quad &|\Phi_2\rangle &= \frac{1}{a} \left( -\frac{\Delta + \alpha}{2 V_{12}} |01\rangle + |10\rangle \right), \nonumber \\
    \epsilon_3 &= \frac{\alpha}{2}, \quad &|\Phi_3\rangle &= \frac{1}{b} \left( \frac{-\nu_1 + \nu_2 + \alpha}{2 V_{12}} |01\rangle + |10\rangle \right), \nonumber \\
    \epsilon_4 &=  \nu_0, \quad &|\Phi_4\rangle &= |11\rangle, \label{eq:eigenstates}
\end{align}
with $\alpha =  \sqrt{\Delta^2 + 4 V_{12}^2}$, and normalization constants $a$, $b$ ensuring unit norm. In the resonant case $\nu_1 = \nu_2$, the central eigenstates reduce to symmetric and antisymmetric Bell states.We assume the dimer is initially in the Gibbs state of the full Hamiltonian $H_{\text{dimer}}$, representing global thermal equilibrium:
\begin{equation}
\eta_{\text{th}}(0) = \frac{1}{\mathcal{Z}} e^{-H_{\text{dimer}} / T}, \quad
\mathcal{Z} = 2 \left[ \cosh\left(\frac{\nu_1 + \nu_2}{2 T}\right) + \cosh\left(\frac{\alpha}{2 T}\right) \right], \label{eq:Gibbs}
\end{equation}
Physically, this corresponds to the two sites being coupled to a common thermal reservoir (e.g., the vibrational environment of the host matrix), which equilibrates the interacting system as a whole. Organic dimers typically reach this equilibrium over timescales \cite{baz2018thermodynamic,li2019timescales} ensuring that the Gibbs distribution is well established. The corresponding density matrix in the computational basis is:
\begin{equation}
    \eta_{\text{th}}(0) =
    \begin{bmatrix}
        \eta_{th_{11}} & 0 & 0 & 0 \\
        0 & \eta_{th_{22}} & \eta_{th_{23}} & 0 \\
        0 & \eta_{th_{32}} & \eta_{th_{33}} & 0 \\
        0 & 0 & 0 & \eta_{th_{44}}
    \end{bmatrix}, \label{eq:rho_matrix}
\end{equation}
with coherence $\eta_{th_{23}} = \eta_{th_{32}}^*$ quantifying the off-diagonal contributions in the single-excitation manifold. The matrix elements are given by:
\begin{align}
    \eta_{th_{11}} &= \frac{1}{\mathcal{Z}} e^{ \nu_0 / T}, \nonumber \\
    \eta_{th_{22}} &= \frac{1}{\mathcal{Z} \alpha} \left[  \Delta \sinh\left(\frac{\alpha}{2 T}\right) + \alpha \cosh\left(\frac{\alpha}{2 T}\right) \right], \nonumber \\
    \eta_{th_{33}} &= \frac{1}{\mathcal{Z\alpha}}\left[(-\nu_1+\nu_2)\sinh \left(\frac{\alpha}{2  T}\right)+\alpha\cosh \left(\frac{\alpha}{2 T}\right)\right], \nonumber \\
    \eta_{th_{23}} &= -\frac{2 V_{12} \sinh(\frac{\alpha}{2 T})}{\alpha \mathcal{Z}}, \nonumber \\
    \eta_{th_{44}} &= \frac{1}{\mathcal{Z}} e^{- \nu_0 / T}. \label{eq:rho_elements}
\end{align}
This thermal state will serve as the initial condition for coherent charging under unitary evolution, which we explore in the following section.

\subsection{Unitary charging dynamics}
We consider an array of dimeric PBI molecules as a QB, where energy is injected through a time-dependent charging protocol. In our model, the PBI–QB is treated as an isolated two-emitter system, so that all charging operations follow unitary dynamics. It is well known that for any non-passive quantum state there exists a unitary transformation that increases its mean energy with respect to the battery Hamiltonian $H_{\text{dimer}}$; the only states that cannot be further charged are passive states, which are diagonal in the eigenbasis of $H_{\text{dimer}}$ with populations arranged in non-increasing order of energy \cite{allahverdyan2004maximal,lenard1978thermodynamical,pusz1978passive,salvia2021distribution}.

\begin{figure}[ht]
    \centering
    \includegraphics[width=0.4\linewidth]{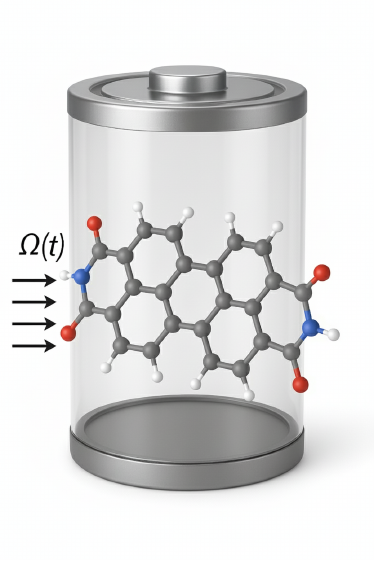}
    \caption{Schematic of the PBI-QB charging protocol. The PBI molecules, initialized in a Gibbs state at thermal equilibrium, are charged via a local field pulse $\Omega(t)=\Omega$ supplying the required energy. }
    \label{fig02}
\end{figure}

In our charging protocol, we apply a coherent drive along the $x$–axis with amplitude $\Omega(t)=\Omega$ (see Fig. \ref{fig02}). Physically, this corresponds to the action of an ultrafast optical pulse that induces Rabi oscillations between the ground and excited states of each PBI monomer, thus implementing an effective $X$-rotation in the two-level basis \cite{julia2020bounds,ghosh2022dimensional,ghosh2020enhancement}. Typical femtosecond pulses can generate Rabi frequencies in the meV range, well within the experimentally accessible regime \cite{genkin2001rabi,rezai2019detuning}. The charging Hamiltonian takes the form
\begin{equation}
H_{\mathrm{ch}} = \Omega \,(\hat{\sigma}_{1}^{x}\otimes \hat{\mathbb{I}}_{2} + \hat{\mathbb{I}}_{1}\otimes \hat{\sigma}_{2}^{x}),
\end{equation}
where $\Omega$ denotes the effective Rabi frequency, taken uniform for simplicity. Once the system approaches its maximum stored energy, the field can be inverted to prevent cyclic return to the initial state. 

This charging operation is formally described by the unitary
\begin{equation}\label{eq:Ux}
U_{X}(t)=\exp\!\left[-\,\mathrm{i}\!\int_{0}^{t}H_{\mathrm{ch}}(t')\,dt'\right]
=\exp\!\left[-\,\mathrm{i}H_{\mathrm{ch}}t\right],
\end{equation}
which, for constant $\Omega$, reduces in the computational basis $\{\lvert00\rangle,\lvert01\rangle,\lvert10\rangle,\lvert11\rangle\}$ to
\begin{equation}\label{eq:Umatrix}
  U_{X}(t) \;=\;
  \begin{pmatrix}
    \phi & -\eta_- & -\eta_- & -\vartheta\\
   -\eta_- & \phi & -\vartheta & -\eta_-\\
   -\eta_- & -\vartheta & \phi & -\eta_-\\
   -\vartheta & -\eta_- & -\eta_- & \phi
  \end{pmatrix},
\end{equation}
with parameters
\[
\phi=\cos^2(\tau),\qquad \eta_-=-\sin\tau\cos\tau,\qquad \vartheta=\sin^2(\tau),\qquad \tau=\Omega t.
\]

This protocol therefore realizes a coherent charging of the PBI–QB in a closed setting, in direct analogy with X–gate schemes implemented in spin-chain and superconducting QBs \cite{quach2020using,HaddadiQB2024,elghaayda2025,ali2024ergotropy}.

\section{Performance metrics\label{sec3}}
To evaluate the performance of the PBI-QB, we analyze a full thermodynamic cycle consisting of three stages: initialization in the uncharged state, energy injection during the charging step, and subsequent work extraction. The state of the dimer battery at time $\tau$ is denoted by $\eta(\tau)$ and evolves under the Hamiltonian $H_{\text{dimer}}$. The central objective is to quantify the amount of energy that can be stored and later retrieved under cyclic unitary dynamics. A cyclic operation requires that the Hamiltonian returns to its original form at the beginning and end of the protocol, $H_{\text{dimer}}(0) = H_{\text{dimer}}(\tau) = H_{\text{dimer}}$~\cite{binder2015quantacell}. Because the dynamics are unitary, variations in the expectation value $\langle H_{\text{dimer}} \rangle$ correspond directly to the work exchanged with the battery.

Within this cycle, the storage capacity establishes the maximum energy that can be theoretically accommodated in the dimer system, serving as a thermodynamic ceiling. The ergotropy quantifies the fraction of this stored energy that is actually extractable through optimal unitary operations. The charging power characterizes the rate at which ergotropy is accumulated during the charging stage, identifying the operational speed of the battery. Finally, quantum coherence captures the non-classical correlations that emerge during the cycle and act as an enabling resource, directly influencing both ergotropy and power. Together, these four metrics provide a comprehensive framework for evaluating the performance of the PBI dimer battery under realistic operating conditions. These indicators have proven fundamental in recent theoretical investigations of spin-chain, cavity-coupled, and molecular QBs \cite{allahverdyan2004maximal,lenard1978thermodynamical,pusz1978passive,binder2015quantacell,ghosh2020enhancement,baumgratz2014quantifying}. Our approach begins with the spectral decomposition of the battery Hamiltonian, expressed as
\begin{equation}\label{spec}
  H_{\text{dimer}} = \sum_{i=1}^{4} \epsilon_i \, |\Phi_i\rangle \langle \Phi_i|, 
  \quad \text{with} \quad \epsilon_{i+1} \geq \epsilon_i,
\end{equation}
where $\{ \epsilon_i \}$ denote the ordered eigenenergies and $\{ |\Phi_i\rangle \}$ the corresponding orthonormal eigenstates. This diagonal form facilitates precise tracking of internal energy shifts and provides a natural basis for defining passive and active states under unitary dynamics. Analogously, the quantum state of the battery is represented in its eigenbasis as
\begin{equation}\label{state}
  \eta(t) = \sum_{j=1}^{4} \lambda_j \, |e_j\rangle \langle e_j|, 
  \quad \text{with} \quad \lambda_{j+1} \leq \lambda_j,
\end{equation}
where the coefficients $\{ \lambda_j \}$ are the eigenvalues of $\eta(t)$, arranged in descending order. 
At $t = 0$, we assume the QB begins in the Gibbs thermal state $\eta_{\text{th}}(0)$ expressed in Eq. \eqref{eq:rho_matrix}, which serves as our reference uncharged and passive QB state~\cite{binder2015quantacell,bakhshinezhad2024trade}. The thermal (Gibbs) state is a special type of passive state: it is diagonal in the energy eigenbasis, with populations decreasing exponentially with increasing energy. As such, it represents a completely passive configuration from which no work can be extracted via unitary operations. Importantly, while all thermal states are passive, not all passive states are thermal \cite{koukoulekidis2021geometry,campaioli2019quantum}.

During the charging phase, we apply a charging unitary operator $\hat{U}_{\text{ch}}$ to transform the QB from its initial passive state $\eta_{\text{th}}(0)$ to an active state $\sigma$. The active state is constructed by pairing the largest eigenvalues $\lambda_j$ of the current state with the highest energy eigenstates $|\Phi_j\rangle$ of the Hamiltonian:
\begin{equation}
\sigma = \sum_{j} \lambda_j |\Phi_j\rangle \langle \Phi_j|, \quad \text{where } \lambda_{j+1} \geq \lambda_j \text{ for all } j.
\label{eq:active_state}
\end{equation}
The energy injection required to achieve this active state is quantified by the anti-ergotropy \cite{yadav2025thermo}, which represents the maximum amount of energy that can be injected via cyclic unitary processes \cite{yang2023battery,yadav2025thermo}
\begin{equation}
\mathcal{W}(t) = \text{Tr}\left[(\eta(t) - \sigma) H_{\text{dimer}}\right].
\label{eq:anti_ergotropy}
\end{equation}
Once the QB is fully charged to its active state, we can extract work from it until it reaches a passive state. 
\subsection{Ergotropy}
The ergotropy $\mathcal{E}(t)$ quantifies the maximal work extractable from a system via any cyclic unitary operation, equivalently measuring the deviation of $\eta$ from its passive counterpart $\eta_{\rm pas} = \sum_i \lambda_i |\Phi_i\rangle \langle \Phi_i|$:
\begin{equation}\label{ergo}
  \mathcal{E}(t)
  = \mathrm{Tr}\bigl[(\eta - \eta_{\rm pas})\,H_{\text{dimer}} \bigr]
  = \sum_{m,n} \lambda_m\,\epsilon_n \bigl( |\langle \Phi_n|e_m\rangle|^2 - \delta_{mn} \bigr).
\end{equation}  
This quantity captures coherence-enhanced work extraction in spin-chain and molecular QB models \cite{ali2024ergotropy,li2025collective}. For clarity, both the ergotropy $\mathcal{E}(t)$ and the anti-ergotropy $\mathcal{W}(t)$ are non-negative: $\mathcal{E}(t)$ measures the extractable work, while $\mathcal{W}(t)$ quantifies the energetic cost to reach the fully active state
\[
\sigma = \sum_j \lambda_j |\Phi_j\rangle \langle \Phi_j|, \quad \text{with } \lambda_{j+1} \geq \lambda_j \text{ for all } j.
\]  
Although direct measurement of ergotropy in molecular systems is challenging, operational estimates can be obtained via spectroscopic population readout (e.g., single-molecule fluorescence, transient absorption, 2D electronic spectroscopy), partial quantum state tomography, calorimetric detection, or coupling to ancillary modes (cavity or vibrational), with signatures such as population inversion, coherent oscillations, or energy transfer to a probe \cite{niu2024experimental,li2025experimental,malavazi2024two}.
\subsection{Charging Power}
Instantaneous power is defined as the rate of change of ergotropy,
\begin{equation}
  \mathcal{P}(t) \;=\; \frac{d\,\mathcal{E}(t)}{dt}.
\end{equation}
The average power up to time $t$,
$\langle p(t)\rangle=\mathcal{E}(t)/t$,
and its maximum over $t$, have been used to benchmark collective charging advantages in Dicke‐type and dark‐state protocols \cite{quach2020using,ferraro2018high}.
\subsection{Capacity}
The capacity of the QB is defined as the difference between the ergotropy and anti-ergotropy~\cite{yang2023battery,yadav2025thermo}:
\begin{equation}
\mathcal{C} = \mathcal{E}(t) - \mathcal{W}(t).
\label{eq:capacity}
\end{equation}
This quantity establishes the fundamental energy window within which the QB can operate during a complete thermodynamic cycle. While both ergotropy and anti-ergotropy may fluctuate dynamically throughout an isentropic evolution, their difference remains invariant under any unitary transformation. As such, the capacity $\mathcal{C}$ provides a robust, state-independent measure of the maximum usable energy span of the battery, reflecting the ultimate limits of work transfer consistent with reversible unitary dynamics.
\subsection{Quantum Coherence}
To quantify the role of quantum coherence during the charging dynamics, we employ the $\ell_1$–norm of coherence~\cite{baumgratz2014quantifying}. The time-evolved state under the X–gate charging protocol is given by
\begin{equation}\label{RX}
\eta(t) = U_{X}(t)~\eta_{\rm th}(0)~U_{X}^\dagger(t),
\end{equation}
from which the coherence is computed as
\begin{equation}\label{eq31}
\mathcal{C}_{\ell_1}[\eta(t)] = \sum_{i\neq j} \bigl|\langle \Phi_i | \eta(t) | \Phi_j\rangle \bigr|,
\end{equation}
where the summation extends over all off-diagonal elements of the density matrix in the energy eigenbasis. This quantity measures the extent of quantum superposition in the system and directly captures nonclassical correlations responsible for enhanced performance beyond classical limits. Furthermore, $\mathcal{C}_{\ell_1}$ has been shown to correlate strongly with extractable work in coherent charging schemes and, in the case of molecular dimers, highlights the role of dipole–dipole interactions in maintaining coherence at finite temperature~\cite{scholes2017using,huelga2013vibrations}.

The quantities $\mathcal{E}$, $\mathcal{P}$, $\mathcal{C}$, and $\mathcal{C}_{\ell_1}$ provide a unified framework to evaluate the PBI–QB and its robustness to realistic conditions. All results are presented in dimensionless units, so that parameters are expressed relative to the molecular transition frequency, highlighting universal trends independent of specific units.

\section{Results \label{sec4}}
\begin{figure}[ht]
\centering
\subfigure[]{\label{figure1a}\includegraphics[scale=0.6]{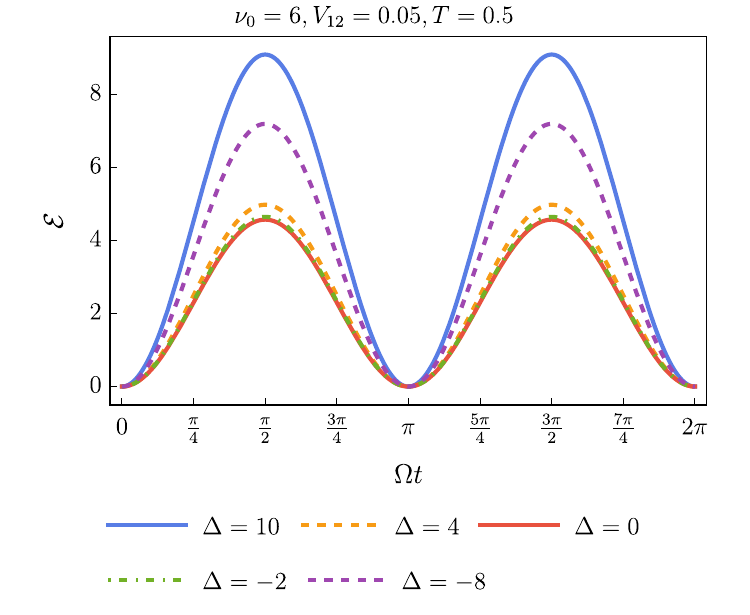}}
\subfigure[]{\label{figure1b}\includegraphics[scale=0.6]{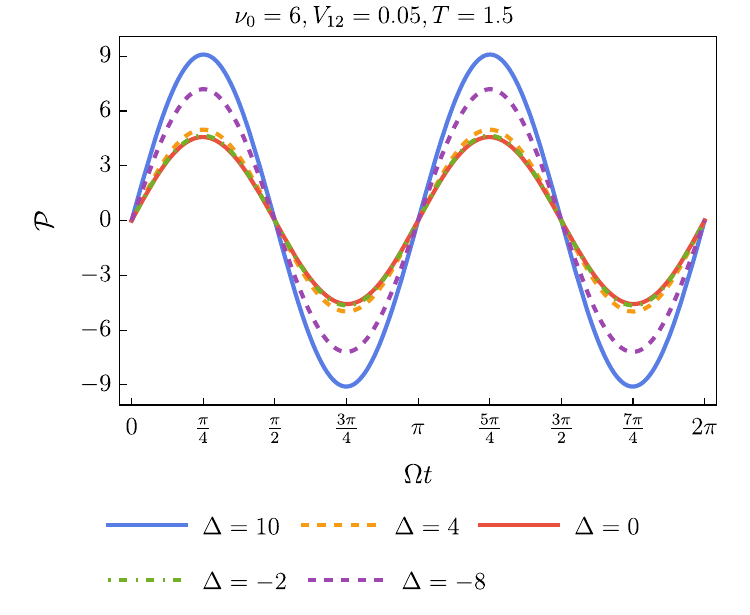}}
\subfigure[]{\label{figure1c}\includegraphics[scale=0.6]{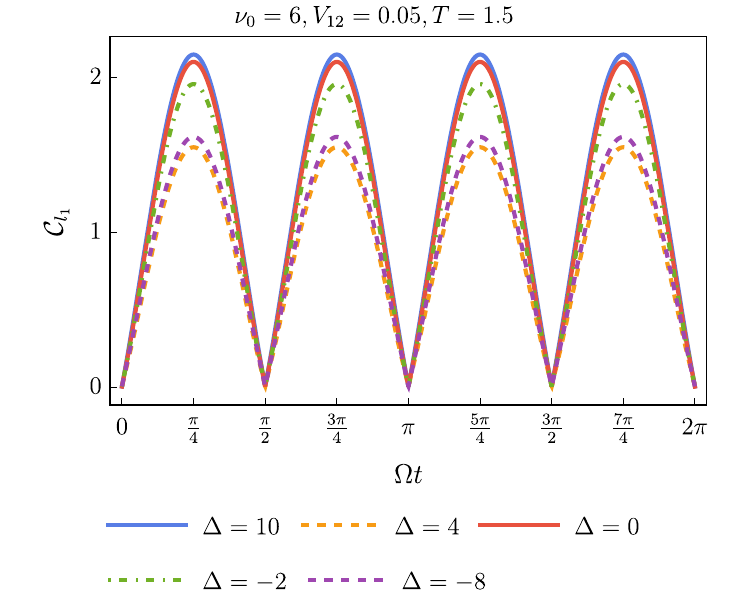}}
\subfigure[]{\label{figure1d}\includegraphics[scale=0.6]{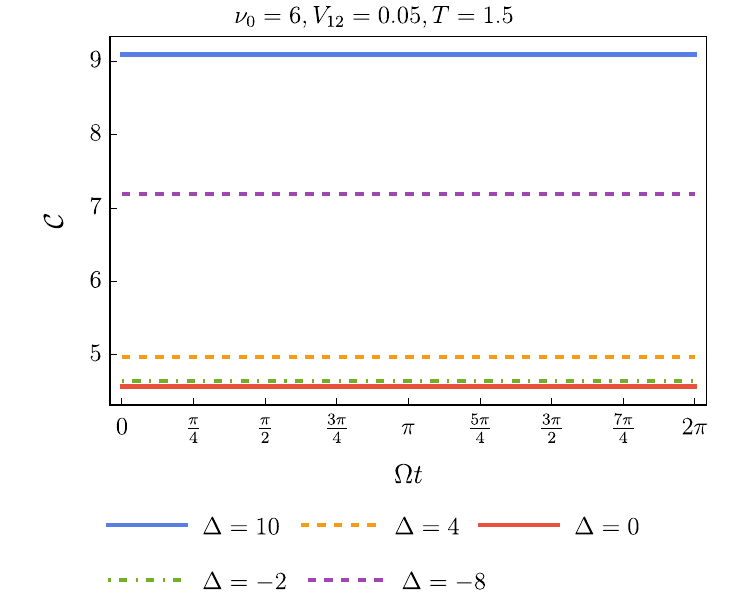}}
\caption{Dynamics of ergotropy $\mathcal{E}$ \ref{figure1a}, instantaneous power $\mathcal{P}$ \ref{figure1b}, coherence $\mathcal{C}_{l_1}$ \ref{figure1c}, and capacity $\mathcal{C}$ \ref{figure1d} for the PBI-QB as functions of $\Omega t$, for $\nu_0=6$, $V_{12}=0.05$, $T=0.5$, under resonant ($\Delta=0$) and detuned ($\Delta\neq 0$) conditions.}
\label{figure1}
\end{figure}

In this section, we investigate the charging dynamics of the PBI-QB through four central performance metrics: ergotropy $\mathcal{E}(\tau)$, instantaneous power $\mathcal{P}(\tau)$, storage capacity $\mathcal{C}$, and the $\ell_1$-norm of coherence $\mathcal{C}_{\ell_1}$. Together, these metrics enable a comprehensive assessment of the extractable work, the rate of energy transfer, the thermodynamic bound on stored energy, and the pivotal role of quantum coherence as a resource underpinning battery performance. Figure~\ref{figure1}, illustrates their comparative behavior under both resonant ($\Delta = 0$) and off-resonant ($\Delta \neq 0$) conditions, thereby highlighting the influence of spectral detuning on the overall charging efficiency and coherence dynamics of the PBI-QB.  

Figure~\ref{figure1} illustrates the key features of the PBI dimeric QB. The ergotropy, shown in Fig.~\ref{figure1a}, exhibits an oscillatory behavior as a function of the driving parameter $\Omega t$ and is strongly influenced by the spectral detuning $\Delta = \nu_1 - \nu_2$. At $\Omega t = k\pi+ \pi/2~ ( k=0,1)$ and finite detuning, the extractable work reaches its maximum, whereas at $\Omega t = k \pi$ it is drastically suppressed. Notably, the peak ergotropy increases with $|\Delta|$ and attains its minimum at exact resonance ($\Delta = 0$). This behavior reflects the fact that spectral symmetry prevents the buildup of population imbalance, whereas controlled detuning introduces asymmetry that enhances the extractable work. The charging power, depicted in Fig.~\ref{figure1b}, follows a qualitatively different trend. It also oscillates with the driving parameter, displaying resonance-like peaks at intermediate values ($\pi/4, 3\pi/4, \dots$). Maxima and minima occur at higher detuning values, identifying optimal operating regimes where energy transfer is accelerated by the spectral mismatch. At exact resonance ($\Delta = 0$), the power is minimized, whereas excessive detuning amplifies the peaks, indicating that asymmetry is crucial to sustain efficient charging. The coherence dynamics, quantified by the $\ell_{1}$-norm in Fig.~\ref{figure1c}, oscillate with the driving field but exhibit a markedly different behavior under detuned regimes. Unlike ergotropy and
power, the coherence displays a non-monotonic dependence on detuning. At exact resonance ($\Delta=0$), coherence is preserved, thereby maintaining the quantum superpositions between the
molecular sites. Interestingly, for strong detuning ($\Delta=10$), coherence is further enhanced, whereas for intermediate values ($\Delta=4$ and $\Delta=-8$), it is suppressed. This highlights the
necessity of carefully fine-tuning the detuning parameter in order to sustain quantum coherence in the system. Overall, this evolution demonstrates that spectral detuning can serve as a powerful control
mechanism to preserve, and even amplify, coherence. Finally, the storage capacity $\mathcal{C}$, reported in Fig.~\ref{figure1d}, sets the thermodynamic ceiling for extractable energy. For a fixed set of parameters $(\nu_0, \Delta, V_{12}, T)$, the capacity $\mathcal{C}$ is independent of the charging time $\Omega t$, as it is invariant under unitary dynamics. Consequently, $\mathcal{C}$ appears as a constant reference line in our plots, while ergotropy and power oscillate in time. Its magnitude, however, remains dependent on $\Delta$: higher detuning values $|\Delta|$ yield elevated capacity plateaus, whereas exact resonance significantly reduces the accessible energy reservoir.

\begin{figure}[ht]
\centering
\subfigure[]{\label{figure2a}\includegraphics[scale=0.6]{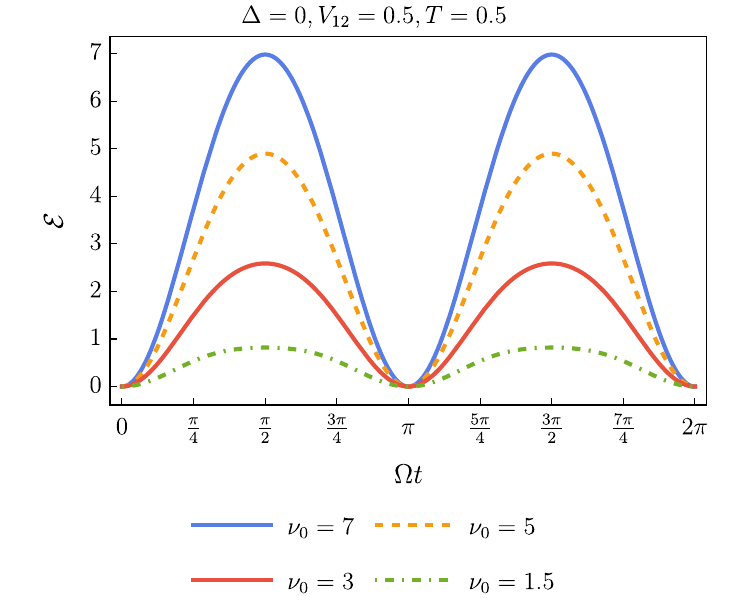}}
\subfigure[]{\label{figure2b}\includegraphics[scale=0.6]{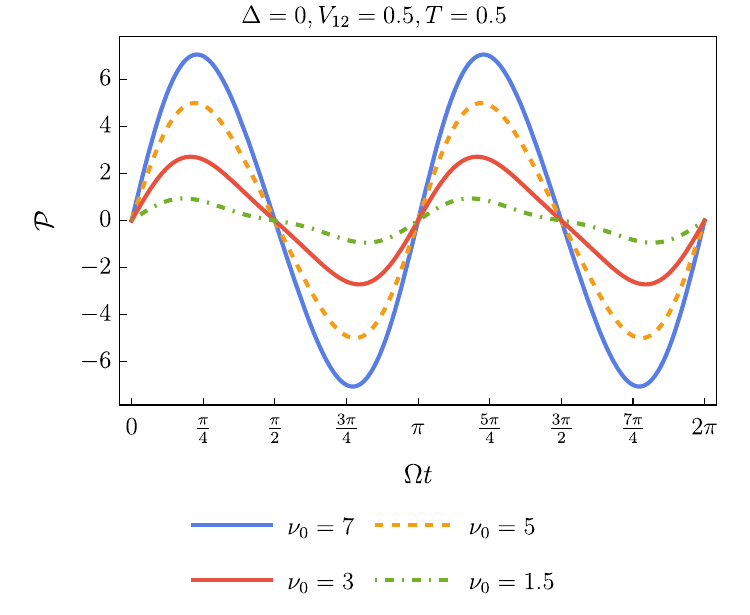}}
\subfigure[]{\label{figure2c}\includegraphics[scale=0.6]{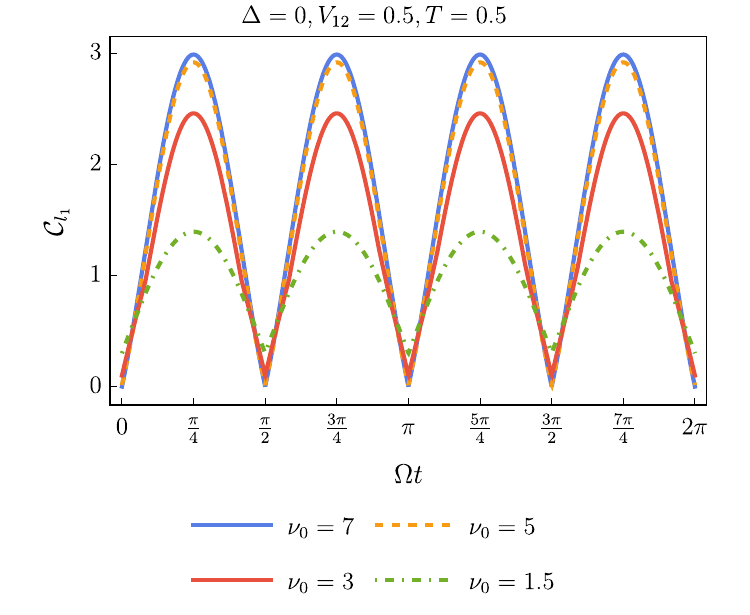}}
\subfigure[]{\label{figure2d}\includegraphics[scale=0.6]{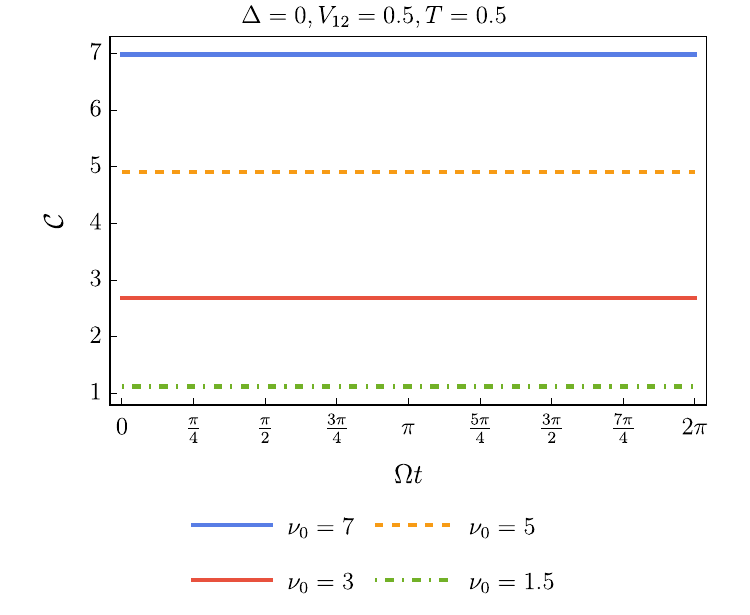}}
\caption{Dynamics of ergotropy $\mathcal{E}$ \ref{figure2a}, instantaneous power $\mathcal{P}$ \ref{figure2b}, coherence $\mathcal{C}_{l_1}$ \ref{figure2c}, and capacity $\mathcal{C}$ \ref{figure2d} for the PBI-QB as a function of $\Omega t$. The parameters are chosen as $\Delta=0$, $V_{12}=0.5$, and $T=0.5$.}
\label{figure2}
\end{figure}
Figure \ref{figure2} shows the impact of the average transition frequency, $\nu_{0} = (\nu_{1}+\nu_{2})/2$, on the charging dynamics of the PBI dimer. Increasing $\nu_{0}$ raises the overall energy scale of the system and proportionally 
enhances all performance indicators. The maximum extractable work 
$\mathcal{E}(\tau)$ increases in amplitude, while its oscillation period remains essentially constant (Fig.~\ref{figure2a}). The instantaneous power $\mathcal{P}(\tau)$ exhibits more pronounced positive peaks as the transition
frequency $\nu_{0}$ increases, reflecting a stronger and more sustained energy absorption. Furthermore,
the positions of these peaks are shifted to later times with increasing $\nu_{0}$, while the overall
oscillation period remains invariant. In parallel, the negative troughs deepen at larger $\nu_{0}$,
signaling a more significant energy backflow during the discharge stage
(Fig.~\ref{figure2b}).
 The $\ell_1$-norm of coherence $\mathcal{C}_{l_1}(\tau)$ exhibits a periodic pattern, peaking at $\tau = \pi/4$ and $\tau = 3\pi/4$ and vanishing at $\tau = \pi/2$. Its amplitude increases with $\nu_0$, reflecting faster generation and revival of coherent superpositions (Fig.~\ref{figure2c}). In parallel, the storage capacity $\mathcal{C}$ increases  with $\nu_{0}$ (Fig.~\ref{figure2d}), since larger transition 
energies expand the accessible spectral range of the dimer battery. This dependence originates from the parameterization of the initial Gibbs ensemble, while the invariance of $\mathcal{C}$ under unitary evolution remains preserved. Overall, these results 
demonstrate that tuning $\nu_{0}$ directly controls the energy scale of the device, simultaneously enhancing ergotropy, charging power, coherence dynamics, and storage capacity.

\begin{figure}[ht]
\centering
\subfigure[]{\label{figure3a}\includegraphics[scale=0.6]{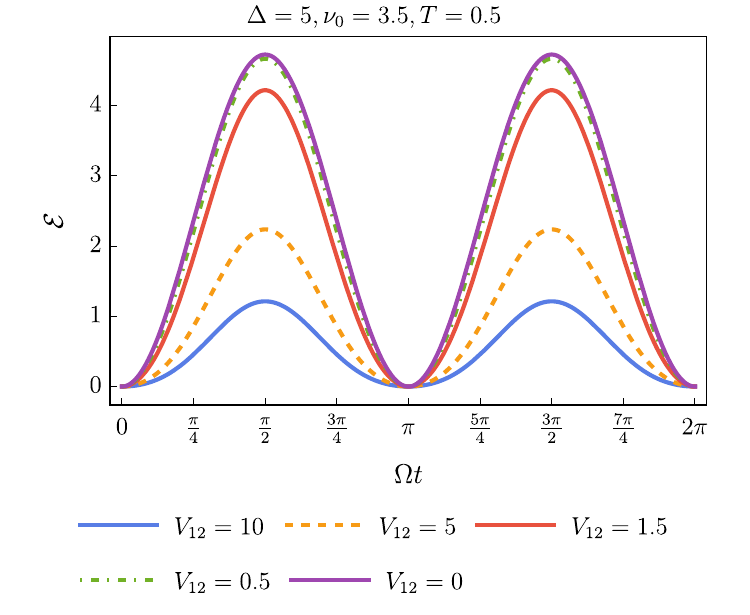}}
\subfigure[]{\label{figure3b}\includegraphics[scale=0.6]{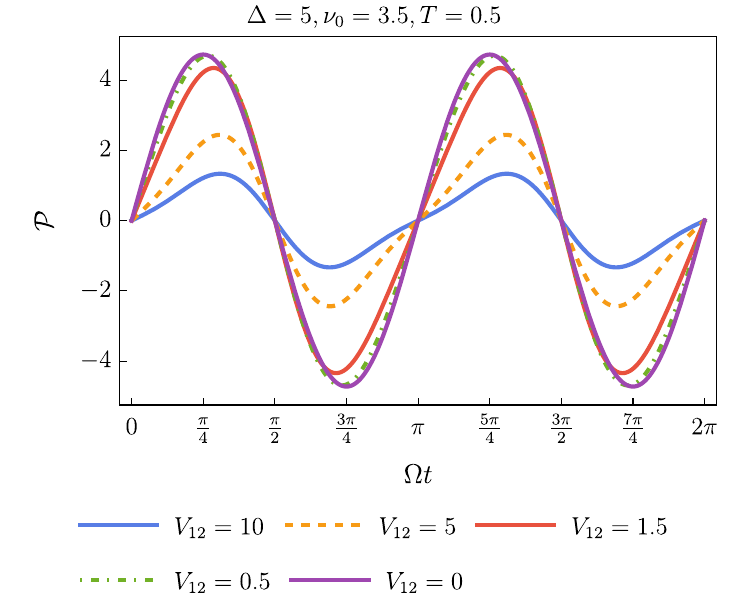}}
\subfigure[]{\label{figure3c}\includegraphics[scale=0.6]{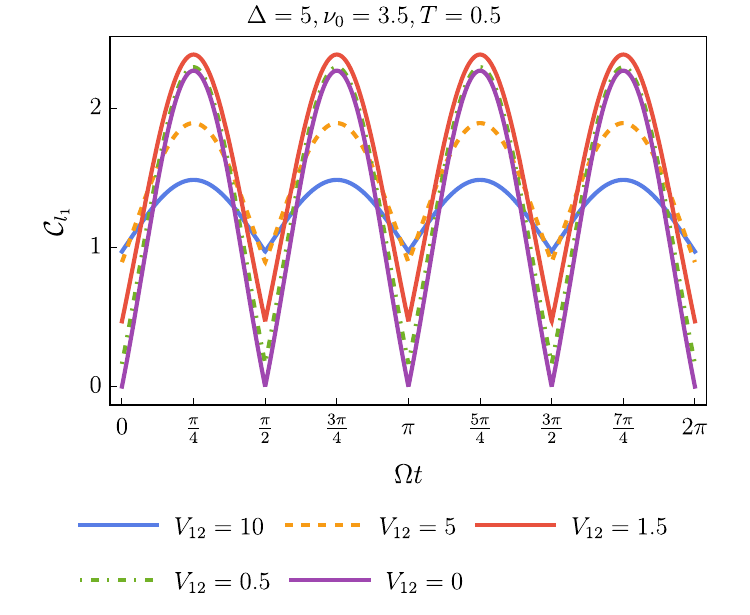}}
\subfigure[]{\label{figure3d}\includegraphics[scale=0.6]{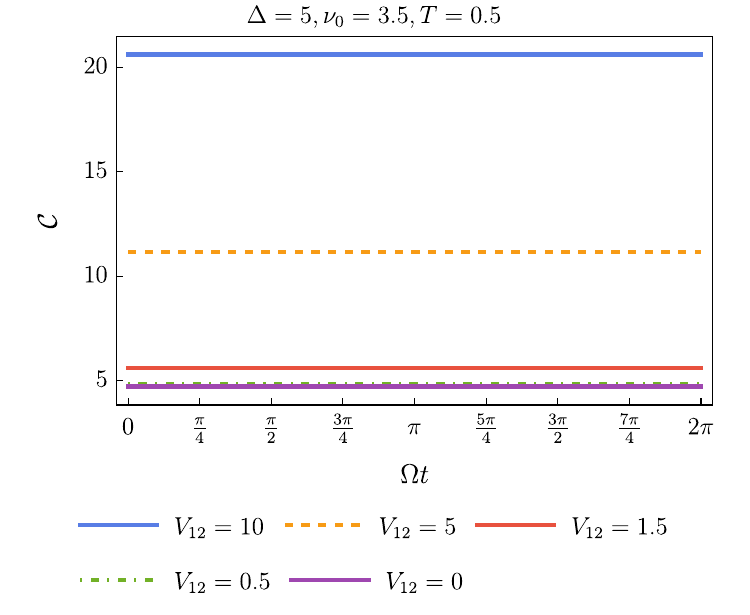}}
\caption{Dynamics of ergotropy $\mathcal{E}$ \ref{figure3a}, instantaneous power $\mathcal{P}$ \ref{figure3b}, coherence $\mathcal{C}_{l_1}$ \ref{figure3c}, and capacity $\mathcal{C}$ \ref{figure3d} for the PBI-QB as a function of $\Omega t$. The parameters are chosen as $\Delta=5$, $\nu_{0}=3.5$, and $T=0.5$.}
\label{figure3}
\end{figure}
Figure \ref{figure3} illustrates the dynamical behavior of the PBI dimer battery as a function of $\Omega t$, for $\Delta=5$, $\nu_0=3.5$, and $T=0.5$, emphasizing the role of the inter-site coupling $V_{12}$. Panel \ref{figure3a} shows that the ergotropy $\mathcal{E}(\tau)$ oscillates with an amplitude that decreases as $V_{12}$ increases, reaching its maximum at $V_{12}=0$. Interestingly, the oscillation frequency and the timing of the first peak remain nearly unaffected by $V_{12}$. In contrast, the instantaneous power $\mathcal{P}(\tau)$ (panel \ref{figure3b}) is more sensitive: while it also peaks at $V_{12}=0$, both its oscillation frequency and the timing of the first maximum shift with increasing $V_{12}$, reflecting a direct influence of coupling on the rate of energy absorption. A distinct behavior emerges for the $\ell_{1}$-norm coherence $\mathcal{C}_{\ell_{1}}(\tau)$ (panel~\ref{figure3c}). While it oscillates in phase with the driving field, the amplitude of these oscillations exhibits a non-regular dependence on the coupling strength $V_{12}$, although the temporal pattern is preserved. Notably, the coherence does not reach its maximum at $V_{12}=0$ but rather at an intermediate value nearest-neighbour electronic coupling  ($V_{12}=1.5$ in our case), whereas for stronger coupling ($V_{12}=10$) it is drastically suppressed. This behavior indicates that properly tuning the interaction strength can enhance coherence, while excessive coupling has the opposite effect. The capacity $\mathcal{C}$ (panel \ref{figure3d}) exhibits the opposite trend: it is minimal when $V_{12}=0$ and increases monotonically with stronger coupling, saturating at large $V_{12}$. This contrasting behavior can be understood as follows: stronger dipolar interactions delocalize excitations across the dimer, spreading populations over a wider energy spectrum, which enlarges the energetic window between passive and active states and thus enhances capacity. At the same time, delocalization reduces the population imbalance that can be coherently extracted, thereby suppressing ergotropy. In the limiting case $V_{12}\to 0$, the dimer behaves as two nearly independent sites, where ergotropy and coher instantaneous power are maximized but capacity is minimal; in the opposite limit $V_{12}\gg\Delta$, the system approaches a  regime with large capacity but very low ergotropy. Overall, these results show that $V_{12}$ plays a dual role: it diminishes ergotropy and power while enhancing the total storable energy. The invariance of the capacity $\mathcal{C}$ under the charging protocol, consistent with its definition as a unitary-invariant resource \cite{yang2023battery}, provides a fixed benchmark against which the efficiency of coherent charging can be assessed.

\begin{figure}[ht]
\centering
\subfigure[]{\label{figure4a}\includegraphics[scale=0.6]{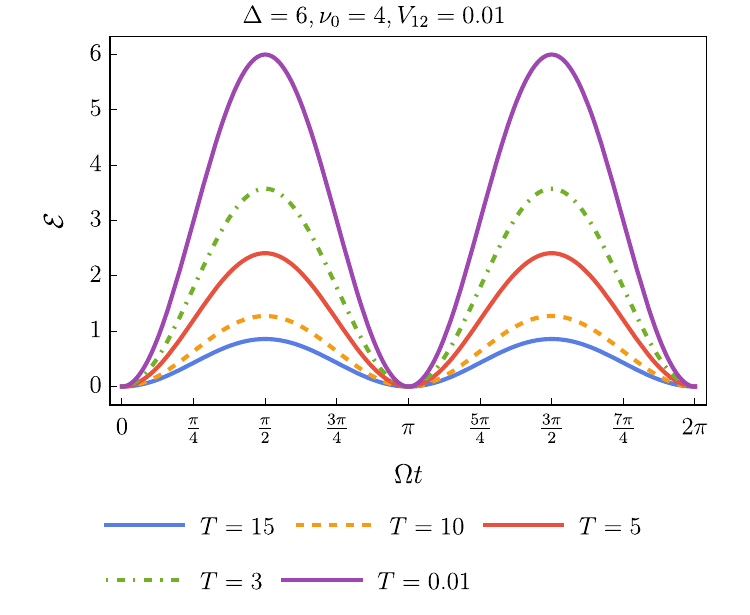}}
\subfigure[]{\label{figure4b}\includegraphics[scale=0.6]{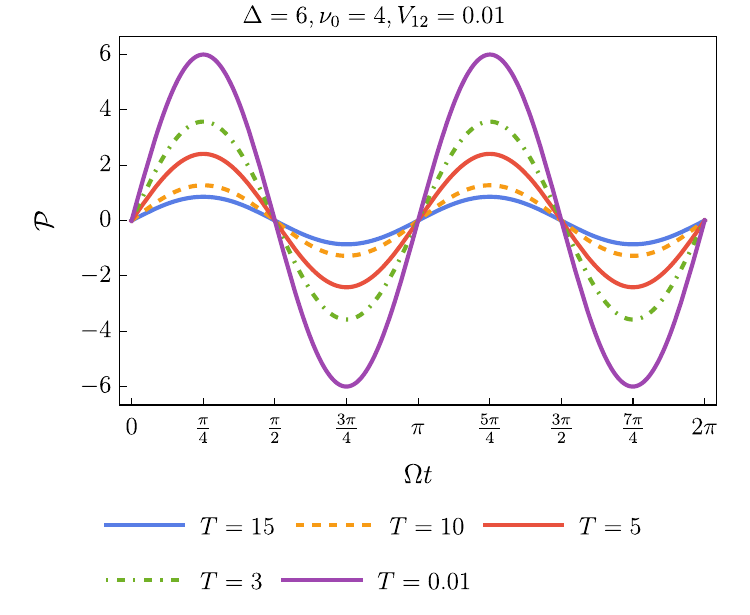}}
\subfigure[]{\label{figure4c}\includegraphics[scale=0.6]{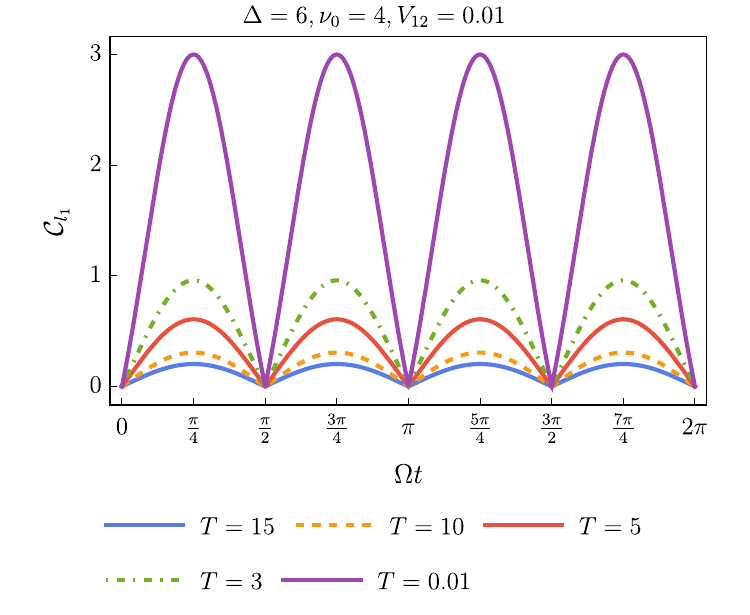}}
\subfigure[]{\label{figure4d}\includegraphics[scale=0.6]{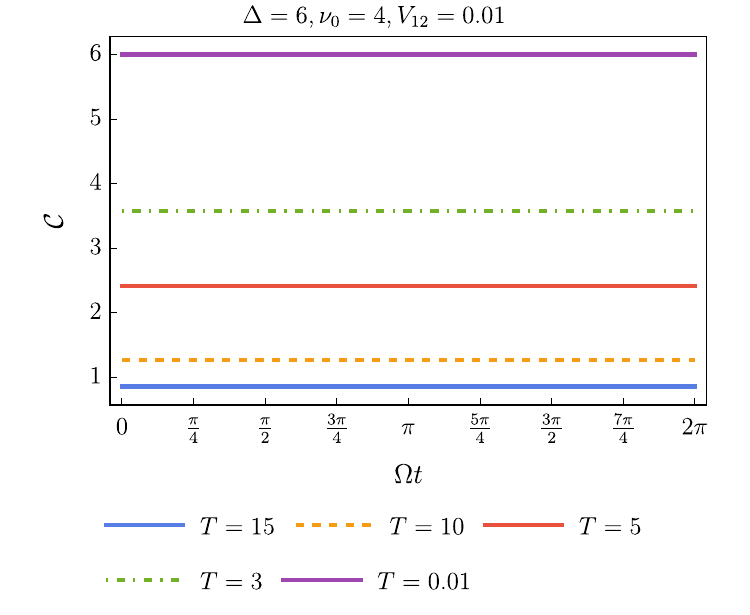}}
\caption{Dynamics of ergotropy $\mathcal{E}$ \ref{figure4a}, instantaneous power $\mathcal{P}$ \ref{figure4b}, coherence $\mathcal{C}_{l_1}$ \ref{figure4c}, and capacity $\mathcal{C}$ \ref{figure4d} for the PBI-QB as a function of $\Omega t$. The parameters are chosen as $\Delta=6$, $\nu_{0}=4$, and $V_{12}=0.01$.}
\label{figure4}
\end{figure}
Figure \ref{figure4} explores the extreme scenario of low dipolar coupling ($V_{12}=0.01$) under significant detuning ($\Delta=6$, $\nu_0=4$) across a range of temperatures $T$. In the absence of inter-site interaction, the system behaves as two independent chromophores, and the peak value of the extractable work $\mathcal{E}(\tau)$ decays rapidly with increasing $T$. This decay occurs because the thermal population of excited states in the Gibbs ensemble reduces the maximum achievable work (see Fig. \ref{figure4a}). The instantaneous power $\mathcal{P}(\tau)$ shown in Fig. \ref{figure4b} reflects this behavior, with its positive charging lobe and negative discharge lobe both shrinking in magnitude and broadening in time at higher temperatures. This indicates that thermal noise lowers peak charging rates but does not prolong the charging/discharging process. The capacity $\mathcal{C}$ remains constant Fig. \ref{figure4d} for a fixed temperature, confirming that the total energy reservoir is unaffected by driving conditions. In Fig.~\ref{figure4c}, the $l_1$-norm of coherence $C_{l_1}(\tau)$, generated solely by the local driving field, exhibits a first peak at $\tau = \pi/4$, vanishes at $\tau = \pi/2$, and then rises to a second peak at $\tau = 3\pi/4$, with this periodic pattern repeating subsequently. However, its amplitude decreases steadily with increasing temperature, indicating that, in the weak-coupling regime, locally generated coherence is highly sensitive to thermal fluctuations. Overall, these trends underscore that while local pulses can induce periodic work extraction and coherence, strong dipolar interactions are essential for harnessing ambient-temperature storage of PBI-QB. 

\section{Conclusion \label{sec5}}
We have demonstrated that dimeric perylene bisimide (PBI) arrays, initialized in a thermal Gibbs state and coherently driven, enable significant work extraction, robust charging power, and sustained quantum coherence under ambient conditions. Through analytical and numerical evaluation of ergotropy, instantaneous power, storage capacity, and the $\ell_1$-norm of coherence, we have shown that exact resonance ($\Delta = 0$) suppresses ergotropy and power due to balanced thermal populations, while finite detuning ($\Delta \neq 0$) breaks this symmetry and substantially enhances both extractable work and charging dynamics. Moreover, the dipole–dipole coupling strength $V_{12}$ plays a critical role in mediating the trade-off between ergotropy, coherence preservation, and storage capacity, with intermediate values optimizing overall performance. We further highlight the role of the average transition frequency $\nu_0$ in setting the energy scale of the system, directly influencing the maximum storable energy and oscillation dynamics. Looking forward, extending this framework to multi-dimer architectures could harness collective phenomena such as superabsorption and quantum synchronization, potentially amplifying charging rates and energy density. Scaling to multi-dimer networks may enable collective superabsorption, where charging power scales superextensively with system size, potentially bridging organic chemistry with macroscopic quantum energy devices. Incorporating open-system effects via Lindblad master equations will be essential to quantify the influence of vibrational dephasing and dissipative losses on performance limits. Future extensions incorporating phonon-assisted dephasing, known to accelerate charging in other QB platforms, could reveal whether vibrational environments play a constructive role in PBI-based storage. These results align with recent advances in the demonstrations of superextensive charging in organic microcavity systems \cite{Quach2022}, controllable charging protocols in NMR-based architectures \cite{Joshi2022}, and the comprehensive perspective provided by the 2024 Colloquium on QBs \cite{campaioli2024colloquium}. Within this broader landscape, PBI dimers emerge as a distinctive organic platform, offering a chemically tunable and scalable complement to superconducting and photonic implementations. Time-resolved ultrafast spectroscopies, already demonstrated on PBI aggregates, provide a natural pathway to experimentally validate ergotropy extraction and coherence lifetimes in fabricated PBI arrays.


\appendix

\section{Density matrix 
$\mathcal{\eta}(t)$ \label{appB}}

The density matrix 
$\mathcal{\eta}(t)$ \eqref{RX} is given by 
\begin{equation}
\mathcal{\eta}(t) = \begin{pmatrix}
\mathcal{\eta}_{11}  & \mathcal{\eta}_{12}  & \mathcal{\eta}_{13}  & \mathcal{\eta}_{14}  \\
\mathcal{\eta}_{21}  & \mathcal{\eta}_{22} & \mathcal{\eta}_{23}  & \mathcal{\eta}_{24}  \\
\mathcal{\eta}_{31}  & \mathcal{\eta}_{32} & \mathcal{\eta}_{33}  & \mathcal{\eta}_{34}  \\
\mathcal{\eta}_{41}  & \mathcal{\eta}_{42}  & \mathcal{\eta}_{43} & \mathcal{\eta}_{44} 
\end{pmatrix},
\end{equation}
where

\small
\begin{equation}
   \mathcal{\eta}_{11}=\frac{2 e^{-\frac{\nu _0}{2 T}} \left(\sin ^4(\tau )+\cos ^4(\tau ) e^{\frac{\nu _0}{T}}\right)+\sin ^2(2 \tau ) \left(\cosh \left(\frac{\alpha
   }{2 T}\right)-\frac{2 V_{12} \sinh \left(\frac{\alpha }{2 T}\right)}{\alpha }\right)}{4 \left(\cosh \left(\frac{\alpha }{2 T}\right)+\cosh
   \left(\frac{\nu _0}{2 T}\right)\right)},
\end{equation}

\small
\begin{equation}
    \mathcal{\eta}_{12}= \frac{\sin (2 \tau ) e^{-\frac{\nu _0}{2 T}} \left(\alpha  \left(\cos (2 \tau )+2 \cos ^2(\tau ) e^{\frac{\nu _0}{T}}-1\right)+2 e^{\frac{\nu
   _0}{2 T}} \left(-\alpha  \cos (2 \tau ) \cosh \left(\frac{\alpha }{2 T}\right)-\sinh \left(\frac{\alpha }{2 T}\right) \left(\Delta -2 V_{12}
   \cos (2 \tau )\right)\right)\right)}{8 \alpha  \left(\cosh \left(\frac{\alpha }{2 T}\right)+\cosh \left(\frac{\nu _0}{2 T}\right)\right)},
\end{equation}

\small
\begin{equation}
    \mathcal{\eta}_{13}=\frac{\sin (2 \tau ) \left(-\cos (2 \tau ) \cosh \left(\frac{\alpha }{2 T}\right)+e^{-\frac{\nu _0}{2 T}} \left(\cos ^2(\tau ) e^{\frac{\nu
   _0}{T}}-\sin ^2(\tau )\right)+\frac{\sinh \left(\frac{\alpha }{2 T}\right) \left(\Delta +2 V_{12} \cos (2 \tau )\right)}{\alpha }\right)}{4
   \left(\cosh \left(\frac{\alpha }{2 T}\right)+\cosh \left(\frac{\nu _0}{2 T}\right)\right)},
\end{equation}

\small
\begin{equation}
   \mathcal{\eta}_{14}=-\frac{\sin ^2(\tau ) \cos ^2(\tau ) \left(\cosh \left(\frac{\alpha }{2 T}\right)-\cosh \left(\frac{\nu _0}{2 T}\right)-\frac{2 V_{12} \sinh
   \left(\frac{\alpha }{2 T}\right)}{\alpha }\right)}{\cosh \left(\frac{\alpha }{2 T}\right)+\cosh \left(\frac{\nu _0}{2 T}\right)},
\end{equation}

\small
\begin{equation}
    \mathcal{\eta}_{22}= \frac{e^{-\frac{\nu _0}{2 T}} \left(\alpha  \sin ^2(2 \tau ) \left(e^{\frac{\nu _0}{T}}+1\right)+\alpha  (\cos (4 \tau )+3) e^{\frac{\nu _0}{2
   T}} \cosh \left(\frac{\alpha }{2 T}\right)+4 e^{\frac{\nu _0}{2 T}} \sinh \left(\frac{\alpha }{2 T}\right) \left(\Delta  \cos (2 \tau
   )+V_{12} \sin ^2(2 \tau )\right)\right)}{8 \alpha  \left(\cosh \left(\frac{\alpha }{2 T}\right)+\cosh \left(\frac{\nu _0}{2 T}\right)\right)},
\end{equation}

\small
\begin{equation}
    \mathcal{\eta}_{23}= \frac{e^{-\frac{\nu _0}{2 T}} \left(\alpha  \sin ^2(2 \tau ) \left(e^{\frac{\nu _0}{T}}+1\right)-8 \alpha  \sin ^2(\tau ) \cos ^2(\tau )
   e^{\frac{\nu _0}{2 T}} \cosh \left(\frac{\alpha }{2 T}\right)-2 V_{12} (\cos (4 \tau )+3) e^{\frac{\nu _0}{2 T}} \sinh \left(\frac{\alpha }{2
   T}\right)\right)}{8 \alpha  \left(\cosh \left(\frac{\alpha }{2 T}\right)+\cosh \left(\frac{\nu _0}{2 T}\right)\right)},
\end{equation}

\small
\begin{equation}
    \mathcal{\eta}_{24}=-\frac{\sin (2 \tau ) \left(-\cos (2 \tau ) \cosh \left(\frac{\alpha }{2 T}\right)+e^{-\frac{\nu _0}{2 T}} \left(\cos ^2(\tau )-\sin ^2(\tau )
   e^{\frac{\nu _0}{T}}\right)+\frac{\sinh \left(\frac{\alpha }{2 T}\right) \left(2 V_{12} \cos (2 \tau )-\Delta \right)}{\alpha }\right)}{4
   \left(\cosh \left(\frac{\alpha }{2 T}\right)+\cosh \left(\frac{\nu _0}{2 T}\right)\right)},
\end{equation}

\small
\begin{equation}
    \mathcal{\eta}_{33}=\frac{e^{-\frac{\nu _0}{2 T}} \left(\alpha  \sin ^2(2 \tau ) \left(e^{\frac{\nu _0}{T}}+1\right)+\alpha  (\cos (4 \tau )+3) e^{\frac{\nu _0}{2
   T}} \cosh \left(\frac{\alpha }{2 T}\right)-2 e^{\frac{\nu _0}{2 T}} \sinh \left(\frac{\alpha }{2 T}\right) \left(2 \Delta  \cos (2 \tau
   )+V_{12} (\cos (4 \tau )-1)\right)\right)}{8 \alpha  \left(\cosh \left(\frac{\alpha }{2 T}\right)+\cosh \left(\frac{\nu _0}{2
   T}\right)\right)},
\end{equation}

\small
\begin{equation}
    \mathcal{\eta}_{34}=-\frac{\sin (2 \tau ) \left(-\cos (2 \tau ) \cosh \left(\frac{\alpha }{2 T}\right)+e^{-\frac{\nu _0}{2 T}} \left(\cos ^2(\tau )-\sin ^2(\tau )
   e^{\frac{\nu _0}{T}}\right)+\frac{\sinh \left(\frac{\alpha }{2 T}\right) \left(\Delta +2 V_{12} \cos (2 \tau )\right)}{\alpha }\right)}{4
   \left(\cosh \left(\frac{\alpha }{2 T}\right)+\cosh \left(\frac{\nu _0}{2 T}\right)\right)},
\end{equation}

\small
\begin{equation}
\mathcal{\eta}_{44}=\frac{2 e^{-\frac{\nu _0}{2 T}} \left(\cos ^4(\tau )+\sin ^4(\tau ) e^{\frac{\nu _0}{T}}\right)+\sin ^2(2 \tau ) \left(\cosh \left(\frac{\alpha
   }{2 T}\right)-\frac{2 V_{12} \sinh \left(\frac{\alpha }{2 T}\right)}{\alpha }\right)}{4 \left(\cosh \left(\frac{\alpha }{2 T}\right)+\cosh
   \left(\frac{\nu _0}{2 T}\right)\right)}.
\end{equation}

Expressions satisfy the standard properties of a density matrix, namely
\[
\eta_{ij}(t) = \eta_{ji}^{*}(t), \qquad \mathrm{Tr}\,\eta(t) = 1.
\]

\section*{Disclosures}
The authors declare that they have no known competing financial interests.

\section*{Data availability}
Numerical data supporting the figures are available from the authors on request.

\bibliography{bibliography} 
\bibliographystyle{ieeetr}
\end{document}